\documentclass[conference]{IEEEtran}

\makeatletter
\def\ps@IEEEtitlepagestyle{%
	\def\@oddfoot{\mycopyrightnotice}%
	\def\@evenfoot{}%
}
\def\mycopyrightnotice{%
	{\footnotesize {}\hfill}
	\gdef\mycopyrightnotice{}
}

\usepackage{cite}
\usepackage{amsmath,amssymb,amsfonts}

\usepackage{algorithmic}
\usepackage{subfigure}

\usepackage{graphicx}
\graphicspath{{figures/}}

\newcommand{\be}{\begin{equation}}
\newcommand{\ee}{\end{equation}}

\usepackage[bookmarks=true]{hyperref}
\hypersetup
{
    colorlinks=true, 
    linktoc=all,     
    linkcolor=black, 
    allcolors=black, 
    bookmarksopen=true,
    bookmarksnumbered=true,
    pdfstartview=, 
    pdfremotestartview=, 
}
\usepackage[all]{hypcap}
\usepackage{bookmark}
\usepackage{cleveref}

\def\BibTeX{{\rm B\kern-.05em{\sc i\kern-.025em b}\kern-.08em
    T\kern-.1667em\lower.7ex\hbox{E}\kern-.125emX}}
    
\begin{document}

\title{Using Capsule Networks to Classify Digitally Modulated Signals with Raw I/Q Data}

\author{\IEEEauthorblockN{James A. Latshaw, Dimitrie C. Popescu and John A. Snoap}
\IEEEauthorblockA{ECE Department, Old Dominion University \\
231 Kaufman Hall, Norfolk, VA 23529, USA \\
\{jlats001, dpopescu, jsnoa001\}@odu.edu \vspace{-0.5cm} }
\and
\IEEEauthorblockN{Chad M. Spooner}
\IEEEauthorblockA{NorthWest Research Associates \\
Monterey, CA 93940, USA \\
cmspooner@nwra.com \vspace{-0.5cm} }
}

\maketitle

\begin{abstract}
Machine learning has become a powerful tool for solving problems in various engineering and science areas, including the area of communication systems.  This paper presents the use of capsule networks for classification of digitally modulated signals using the I/Q signal components.  The generalization ability of a trained capsule network to correctly classify the classes of digitally modulated signals that it has been trained to recognize is also studied by using two different datasets that contain similar classes of digitally modulated signals but that have been generated independently.  Results indicate that the capsule networks are able to achieve high classification accuracy.  However, these networks are susceptible to the datashift problem which will be discussed in this paper.   
\end{abstract}


\section{Introduction}\label{sec:Intro}
In recent years, machine learning has emerged as a powerful tool in solving complex engineering problems and has been applied to
diverse areas ranging from image processing and computer vision to speech recognition and internet search engines.  Machine learning
techniques have also been applied to wireless communication systems and networks in various settings that involve the different layers
of wireless networks to solve problems related to signal classification and recognition of digital modulation schemes at the physical
layer \cite{Tim2017, Tim2018, Snoap_CCNC2022}, to resource and mobility management at the data link (MAC) and network layers,
to localization at the application layer  \cite{Sun_etal_CommSurvey2019}.

Blind classification of digitally modulated signals has usually been accomplished using signal processing techniques that include likelihood-based
approaches \cite{Hameed_etal_TW2009,Xu_etal_TSMC2011} or cyclostationary signal processing (CSP) \cite{li2020automatic}. We note that,
unlike the likelihood- or CSP-based approaches, which require the implementation of complex algorithms to extract features that distinguish
different digital modulation schemes and then use these features for signal classification, machine learning implements neural networks and
relies on their extensive training to make the distinction among different classes of digitally modulated signals. In this direction, recent approaches
use convolutional and residual neural networks with the I/Q signal components for training and signal recognition/classification \cite{Tim2018,
Tim2017, Snoap_CCNC2022} or with alternative signal features such as the amplitude/phase or frequency domain representations of digitally
modulated signals \cite{Sun2018, Zhang2020, Kulin2018, Rajendran2018, Bu2020}.

In this paper we explore the use of capsule networks \cite{Sabour_etal_NIPC2017} to classify digitally modulated signals using raw I/Q signal data.
As discussed in \cite{Sabour_etal_NIPC2017}, capsule networks are a version of convolutional neural networks that emphasize learning desirable
characteristics of the training dataset by means of capsules, which are multiple parallel and independent nodes that learn class specific characteristics.
Capsule networks have been used for digital modulation classification in \cite{sang2018application, li2020automatic}, and in our paper we pursue a
variant of capsule networks which classifies with high accuracy modulated signals corresponding to several commonly used digital modulation schemes.

We compare the classification performance of the capsule network to that of the convolutional neural network (CNN) and the residual network
(RESNET) used for classification of digitally modulated signals in \cite{Snoap_CCNC2022}, and similar to \cite{Snoap_CCNC2022}, we also explore
the problem of the dataset shift, also referred to as out-of-distribution generalization \cite{Djolonga_etal_arXiv2020}. This is an important problem
in machine-learning-based approaches, which occurs when the training and testing data sets are distinct, implying that data from the testing
environment is not used for training the classifier. In this direction we use two datasets that are publicly available from \cite{CSPblog_DataSets}:
\vspace{-0.05cm}
\begin{itemize}
\item \texttt{DataSet1}: This dataset was also used in \cite{Snoap_CCNC2022} and includes signals corresponding to eight common digital
modulation schemes. Details about the characteristics of the digitally modulated signals included in \texttt{DataSet1} are provided in
Section~\ref{sec:Datasets}.
\vspace{-0.05cm}
\item \texttt{DataSet2}: This dataset is similar to \texttt{DataSet1} in terms of signal types included and their characteristics, and the key difference
between signals in \texttt{DataSet1} and \texttt{DataSet2} is that they have different carrier frequency offset (CFO) intervals, which will be used to
study the effects of the dataset shift on the capsule network performance.
\end{itemize}
\vspace{-0.05cm}
Additional details about these two data sets will be provided in section \ref{sec:Datasets}.

The paper is organized as follows: Section~\ref{sec:NNModels} includes a brief introduction to capsule networks along with a description of the
specific capsule network used for classification of digitally modulated signals in our work. This is followed by a description of the data sets used
for training the capsule network and for testing its classification performance in Section~\ref{sec:Datasets}, with performance results displayed
in Section \ref{sec:benchmarkTesting}. The paper concludes with final remarks in Section~\ref{sec:Conclusion}.

\section{Capsule Networks for Digital Modulation Classification}\label{sec:NNModels}
\vspace{-0.1cm}
The aim of capsule networks is to focus on learning desirable characteristics of the input pattern or signal, which correspond to a specific
input class, and they have been used in attempts to emulate human vision. We note that, when an eye receives visual stimulus, the eye
does not focus on all available inputs, instead points of fixation are established and these points are used to identify or reconstruct a mental
image of the object of focus \cite{Sabour_etal_NIPC2017}. This specificity is achieved by means of capsules, which are multiple parallel
and independent nodes that learn class specific characteristics and represent points of fixation in the form of a capsule vector. In the
context of classifying digitally modulated signals, the capsules are expected to discover excursion characteristics that are intrinsic to a
modulation type \cite{li2020automatic, sang2018application}.

\begin{figure}[bp]
\centering
\vspace{-0.5cm}
\includegraphics[width=\linewidth]{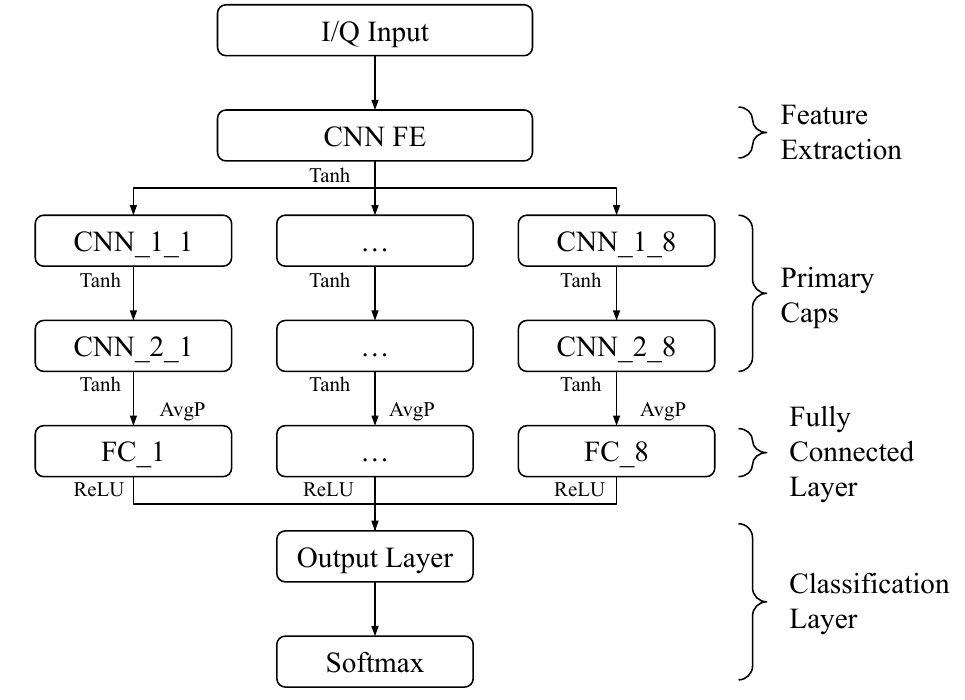}
\vspace{-0.25cm}
\caption{The topology of the considered capsule network for classification of digitally modulated signals with eight branches.}\label{fig:topology}
\vspace{-1cm}
\end{figure}

In general, a capsule network is a shallow convolutional neural network (CNN) that consists of a feature extracting layer followed by parallel
CNN layers referred to as primary caps. Each of the parallel primary cap layers has a ``capsule vector'' as the output, which is referred to as
digit capsule (or digit caps), and has dimension $1 \times N$ neurons. We note that this is different than in the case of CNN approaches, which
rely on a single output neuron per class. The value of $N$ is a design parameter and corresponds to the points of fixation that the capsule
network may learn, which are class specific attributes discovered by the network during training. Ideally, the magnitude of the capsule vector
corresponds to the probability that the input matches the output corresponding to this capsule node and its orientation carries information
related to the input properties. The neurons in the digit caps have connections to neurons in the primary caps layers and can be determined
iteratively using the dynamic routing by agreement algorithm in \cite{Sabour_etal_NIPC2017}, which has also been applied for modulation
classification in  digitally modulated signals \cite{li2020automatic, sang2018application}.

In this paper we consider the capsule network with topology illustrated in Fig.~\ref{fig:topology}, and we study its use in the context of classification
of digitally modulated signals. Furthermore, instead of using dynamic routing by agreement algorithm to update connections between a given capsule 
vector and higher layer neurons, all higher layer neurons are fully connected to each neuron in a $1\times N$ neuron vector. These neurons will
discover desirable attributes in the previous primary caps and activate on these characteristics. We note that, the considered topology allows for
easy implementation from a design perspective and for efficient training as matrix operations are efficiently processed by graphic processing units
(GPUs), whereas iterative learning through dynamic routing by agreement may result in increased computational complexity.

The capsule network shown in Fig.~\ref{fig:topology} takes as input sampled versions of the normalized in-phase (I) and quadrature (Q)
components of a digitally modulated signal, which must be classified as corresponding to one of the following eight digital modulation
schemes: BPSK, QPSK, 8-PSK, DQPSK, MSK, 16-QAM, 64-QAM, and 256-QAM. The various components of the capsule network
considered in Fig.~\ref{fig:topology} include:
\begin{itemize}
\item \texttt{Feature Extraction Layer:} This is the first layer of the network that performs a general feature mapping of the input signal,
and its parameters are inspired from those of the CNNs used for classification of digitally modulated signals in
\cite{Tim2018, Zhou_etal_EURASIP_SP2019, Snoap_CCNC2022}, to include a convolutional layer followed by a batch normalization
layer and an activation function.
\item \texttt{Primary Caps:} This layer consists of a number of primary caps that is equal to the number of digital modulation classes
considered, which are operating in parallel using as inputs the output  from the feature extraction layer. Each primary cap in this layer
includes two convolutional layers with customized filter and stride, and an activation function, and is followed by a fully connected layer.
\item \texttt{Fully Connected Layer:} This layer consists of a $1 \times N$  neuron vector with the weights connecting to the previous
layer. Each neuron in the last layer of the primary caps layer will be fully connected to each neuron in this layer, which are expected
to, ideally, discover characteristics specific to the capsules class. To make the output of the network be compatible with a softmax
classification layer, each neuron within this layer is fully connected to a single output neuron, and the output neurons for all primary
caps will be combined depth wise to produce an \linebreak[4] 8-dimensional vector \textbf{n}, which is passed to the classification layer.
The value of each respective element of \textbf{n} will be representative of the likelihood that its corresponding modulation type is
present in the I/Q input data.
\item \texttt{Classification Layer:} This vector \textbf{n} is passed to the softmax layer, which will map each element $n_i$,
$i=1,\ldots,8$, in \textbf{n} to a value $\sigma_i (\textbf{n})$ that is between $[0,1]$, with each element representing the probability
of occurrence, such that the sum of elements in \textbf{n} adds up to $1$ \cite{luce2008luce}.
\begin{equation}
\sigma_i (\textbf{n}) =  \frac{e^{n_{i}}}{\displaystyle \sum_{j=1}^{8} e^{n_{j}}} \label{eq_sm}
\end{equation}
This provides a convenient way to determine which modulation type is most likely to correspond to the signal with the I/Q data
at the input of the capsule network.
\end{itemize}

More specific details on the capsule network parameters, such as filter sizes, strides, output dimensions, etc., are given in
Table~\ref{table:CNNnetwork}.

\vspace{-0.3cm}
\begin{table}[h]
\centering
\caption{Convolutional Neural Network Layout}
{
\vspace{-0.25cm}
\begin{tabular}{ c c c c}
\hline\hline
Layer & Filter & Stride & Size/Weights \\
\hline
Input & & & 2x32,768\\
Conv & [1,22] & [1,9] & 22x2x64 \\
Batch Normalization		    & $ $ \\
Tanh					        & $ $ \\
Conv-1-(i)	&[1 23]& [1,7] & 23x64x48	 \\
Batch Normalization-1-(i)		& \\
Tanh-1-(i)					    & \\
Conv-2-(i)	& [1 22] & [1,8]	 & 22x48x64 \\
Batch Normalization-2-(i)		& \\
Tanh-2-(i)					    &  \\
Average Pool (i) & [1,8] & [1,1]  &  \\
FC-(i)& &		                & 32 \\
Batch Normalization-3-(i)		& \\
ReLu-1-(i)					    &  \\
Point FC-(i)		   & &       	& 1\\
Depth Concatenation(i=1:8)     & & & 8  \\
SoftMax	                    &   \\
\hline\hline
\end{tabular}
\vspace{-0.25cm}
}\label{table:CNNnetwork}
\vspace{-0.2cm}
\end{table}

\section{Datasets for Capsule Network \\ Training and Testing}\label{sec:Datasets}
A capsule network with the structure outlined in Section~\ref{sec:NNModels}  is trained and tested using digitally modulated signals in two
distinct datasets that are publicly available for general use~\cite{CSPblog_DataSets}. These are referred to as \texttt{DataSet1} and \texttt{DataSet2},
respectively, and each of them contains collections of the I/Q data corresponding to a total of $112,000$ computer generated digitally modulated
signals that include BPSK, QPSK, 8-PSK, DQPSK, 16-QAM, 64-QAM, 256-QAM, and MSK modulation schemes. Signals employ square-root
raised-cosine (SRRC) pulse shaping with roll off factor in the interval $[0.1, 1.0]$ and a total of $32,768$ samples for each signal are included
in the datasets.

We note that the listed signal-to-noise ratios (SNRs) for the signals in both \texttt{DataSet1} and \texttt{DataSet2} correspond to in-band SNR
values, and that a band-of-interest (BOI) detector \cite{BOIdetector} was used to validate the labeled SNRs, CFOs, and SRRC roll-off values
for the signals in both datasets.

\subsection{\texttt{DataSet1}}
\texttt{DataSet1} is available for download as \texttt{CSPB.ML.2018} from \cite{CSPblog_DataSets}, and additional characteristics for signals in this
dataset include:
\begin{itemize}
\item Symbol rates vary between $1$ and $23$ samples/symbol.
\item The in-band SNR varies between $0$ and $12$~dB.
\end{itemize}

\vspace{-0.1cm}
\subsection{\texttt{DataSet2}}
\texttt{DataSet2} is available for download as \texttt{CSPB.ML.2022} from \cite{CSPblog_DataSets}, and for this dataset the additional signal
characteristics include:
\begin{itemize}
\item Symbol rates vary between $1$ and $29$ samples/symbol.
\item The in-band SNR varies between $1$ and $18$~dB. 
\end{itemize}
The symbol rates and CFOs differ between the two datasets, and this difference will be used to study the generalization
ability of the trained capsule networks. We note that the CFOs are distributed over disjoint intervals.

\subsection{Data Augmentation}
It is expected that the trained digital modulation classifier will perform well at classifying high in-band SNR signals and perform less desirably
as the SNR decreases. This is visible in similar digital modulation classification approaches \cite{Snoap_CCNC2022, Tim2017, Tim2018, 
Zhou_etal_EURASIP_SP2019}. However, as noted in \cite{James-thesis}, for the two datasets used there are few samples for the lower
SNR values, and the small sample size for the lower SNR values is not meaningful in evaluating the capsule network performance for the
lower SNR values. To overcome this aspect, a process of data augmentation was used. The process, which is discussed in \cite{James-thesis},
consists of adding random noise to higher SNR signals to reduce the overall in-band SNR and implies after data augmentation the following
in-band ranges:
\begin{itemize}
\item From $-2$~dB to $12$~dB for \texttt{DataSet1}.
\item From $3.5$~dB to $14$~dB for \texttt{DataSet2}.
\end{itemize}

\section{Capsule Network Training \\ and Numerical Results}\label{sec:benchmarkTesting}
\vspace{-0.15cm}
To illustrate the performance of the chosen capsule network for classifying digitally modulated signals we have trained and tested it with
signals from the two datasets available. Each dataset was divided into three subsets that were used for training, validation, and testing,
such that for both \texttt{DataSet1} and \texttt{DataSet2}, 70\% of the data is used for training, 5\% of the data is used for validation and
25\% is used for testing. All data is normalized to unit power before training begins, and the stochastic gradient descent with momentum
(SGDM) algorithm \cite{SGDM} was used for training with a mini-batch size of 250. The capsule networks have been implemented in
MATLAB and trained on a high-performance computing cluster with $18$~NVidia V100 graphical processing unit (GPU) nodes available,
with each node having $128$~GB of memory. We note that training is computationally intensive however, if the resources are leveraged
correctly and the entire dataset is loaded into RAM, training can be completed in several hours.

The classification performance of the capsule network is compared to that of the alternative deep learning approaches for classifying digitally
modulated signals in \cite{Snoap_CCNC2022}, which also use the I/Q signal data but employ a convolutional neural network (CNN) and a
residual network (RESNET) for signal classification. We note that the CNN and RESNET used in \cite{Snoap_CCNC2022} are similar to the neural
network structures considered in~\cite{Tim2018} and yield similar results those in \cite{Tim2018} when tested with signals that come from
the same dataset as the one used for training. However, as discussed in \cite{Snoap_CCNC2022}, these NNs do not display meaningful
generalization ability, and both the CNN and the RESNET fail to identify most digital modulation schemes which they have been trained to recognize
when tested with signals from a dataset that was generated independently from the training one. 

\begin{figure}
\centering
\includegraphics[scale=0.35]{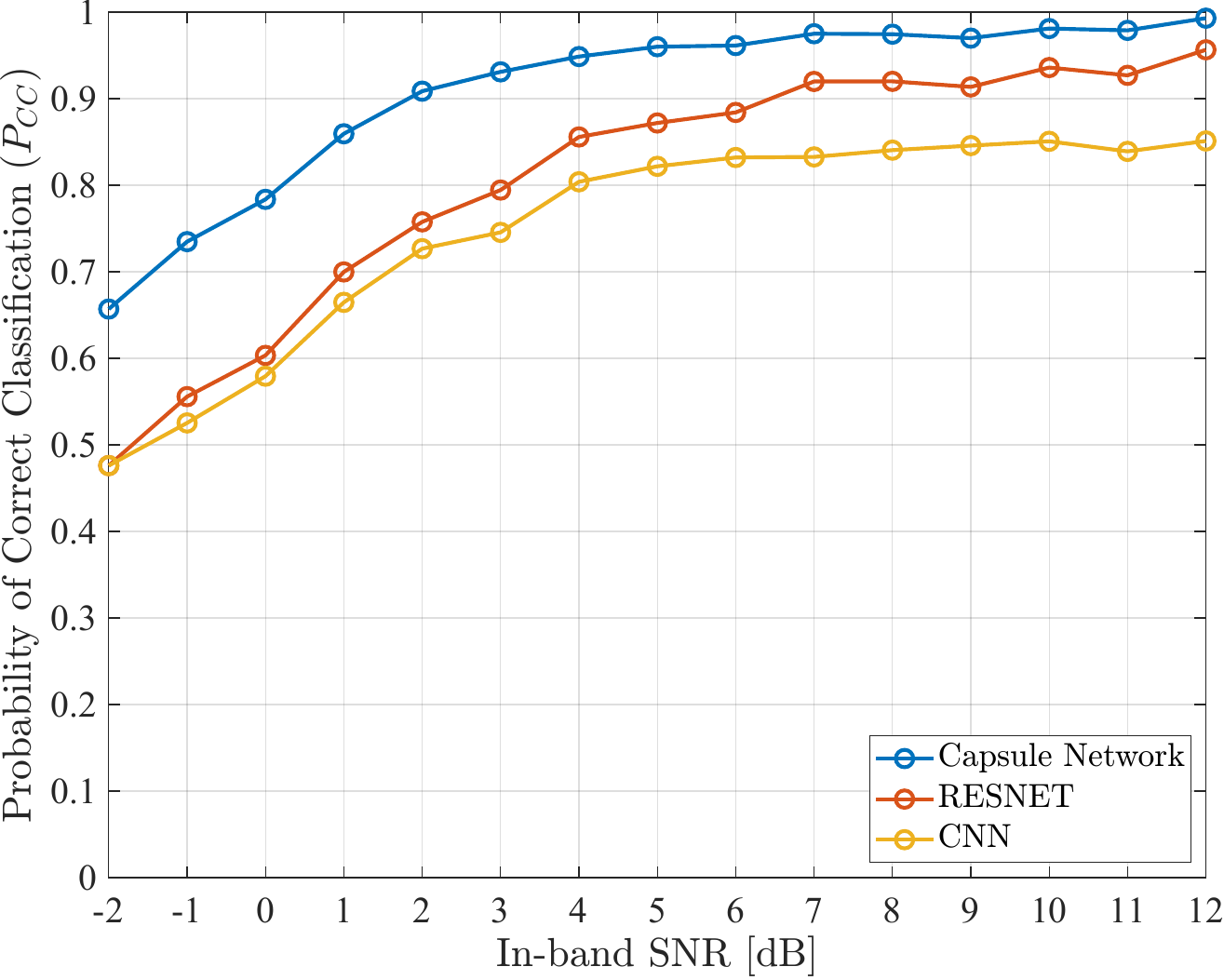}
\vspace{-0.2cm}
\caption{Probability of correct classification versus SNR when both training and test data come from \texttt{DataSet1}.}\label{fig:snr}
\end{figure}

\begin{figure}
\centering
\includegraphics[scale=0.35]{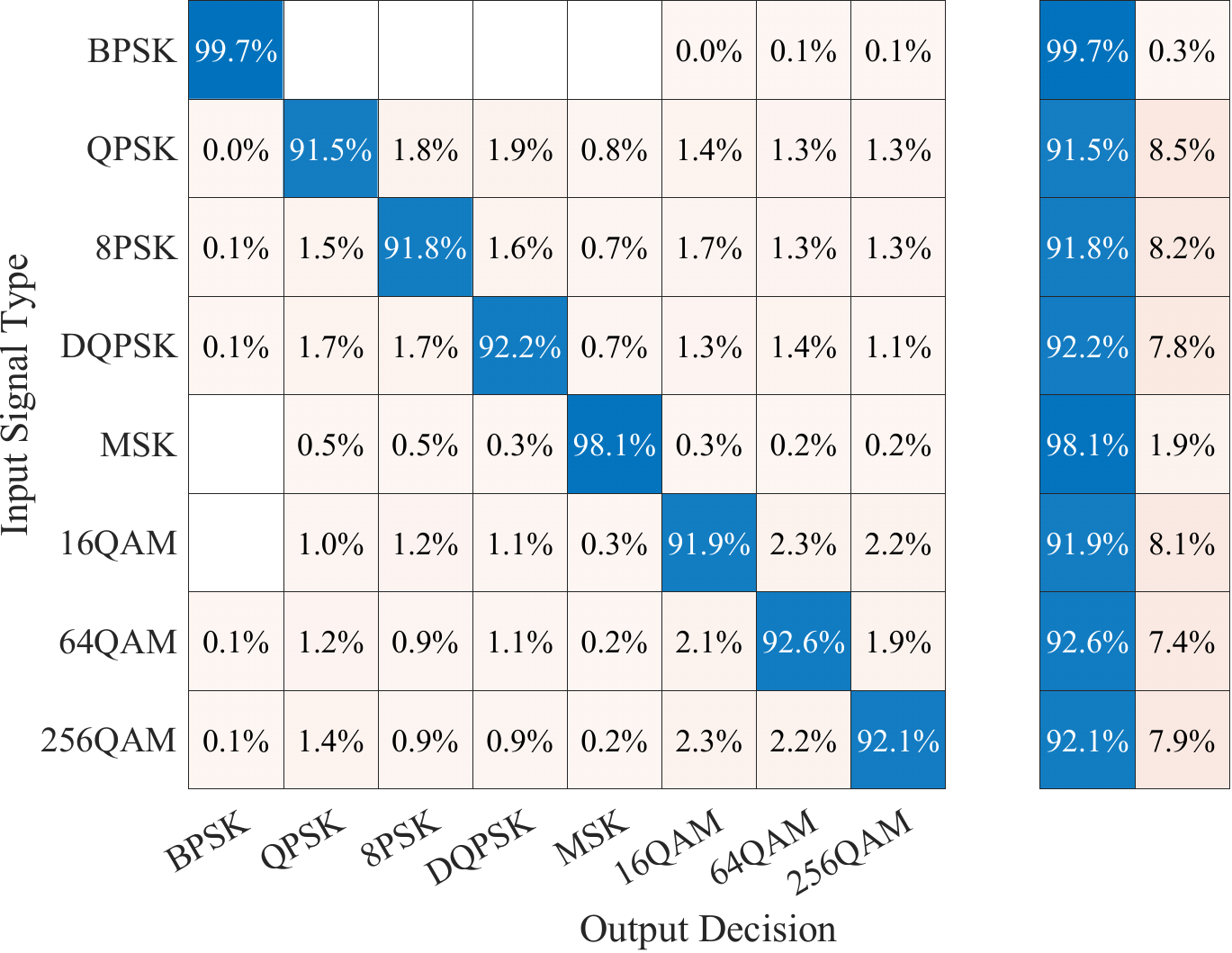}
\vspace{-0.2cm}
\caption{Confusion matrix when both training and test data come from \texttt{DataSet1}.}\label{fig:cm}
\vspace{-0.5cm}
\end{figure}

In the first experiment performed, the capsule network outlined in Fig.~\ref{fig:topology} is trained using \texttt{DataSet1}. Results from
this experiment showing the probability of correct classification and the confusion matrix are given in Fig.~\ref{fig:snr} and Fig.~\ref{fig:cm},
respectively, from which we note an overall performance of $93.7357\%$ correct classification with probabilities of correct classification of
individual modulation schemes ranging from $91.9\%$ for 16-QAM to $99.7\%$ for BPSK. 


\begin{figure}
\centering
\includegraphics[scale=0.35]{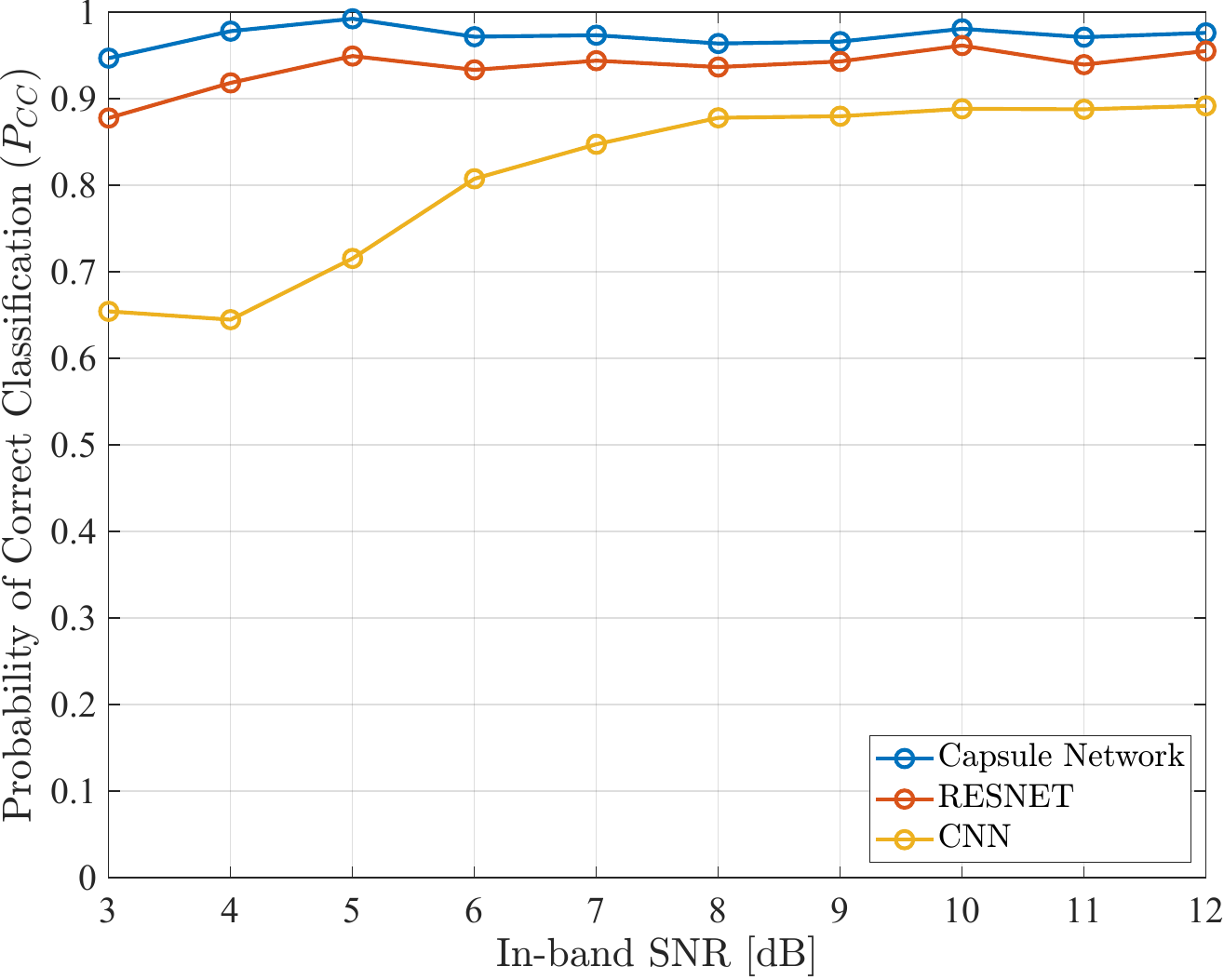}
\vspace{-0.2cm}
\caption{Probability of correct classification versus SNR when both training and test data come from \texttt{DataSet2}.}\label{fig:snr_gc}
\end{figure}

\begin{figure}
\centering
\includegraphics[scale=0.35]{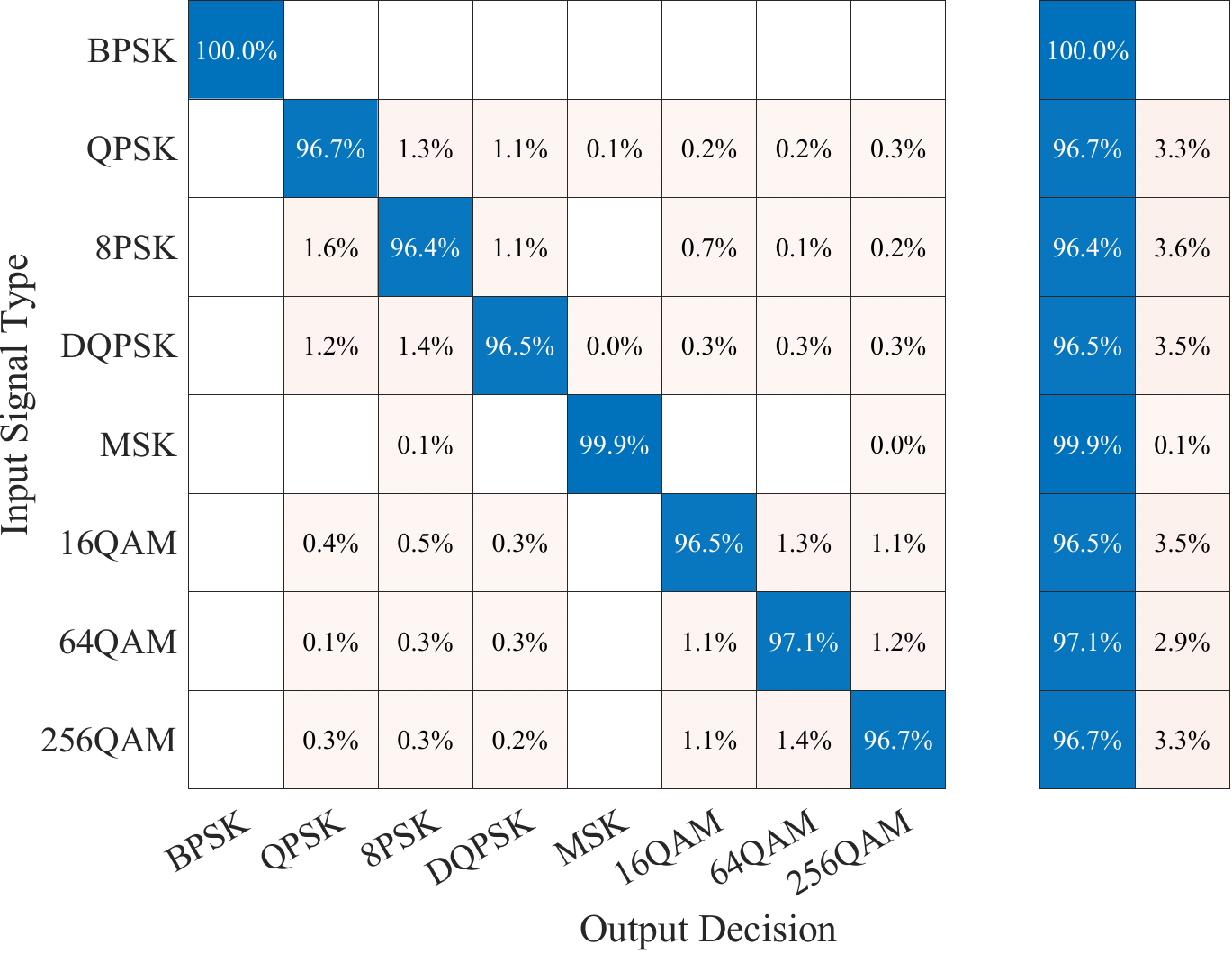}
\vspace{-0.2cm}
\caption{Confusion matrix when both training and test data come from \texttt{DataSet2}.}\label{fig:cm_gc}
\vspace{-0.35cm}
\end{figure}

In a following experiment, the capsule network is re-trained using \texttt{DataSet2}, and similar results are obtained as can be seen
from Fig.~\ref{fig:snr_gc} and Fig.~\ref{fig:cm_gc},  showing the probability of correct classification and the confusion matrix, respectively,
that are achieved when \texttt{DataSet2} is used. In this case we note an overall performance of $97.4607\%$ correct classification with
probabilities of correct classification of individual modulation schemes ranging from $96.4\%$ for 8-PSK to $100\%$ for BPSK. The slight
improvement in classification performance displayed in this case due to the fact that the SNR range for signals in \texttt{DataSet2} is more
favorable than the SNR range for signals in \texttt{DataSet1}.

\begin{figure}
\centering
\includegraphics[scale=0.35]{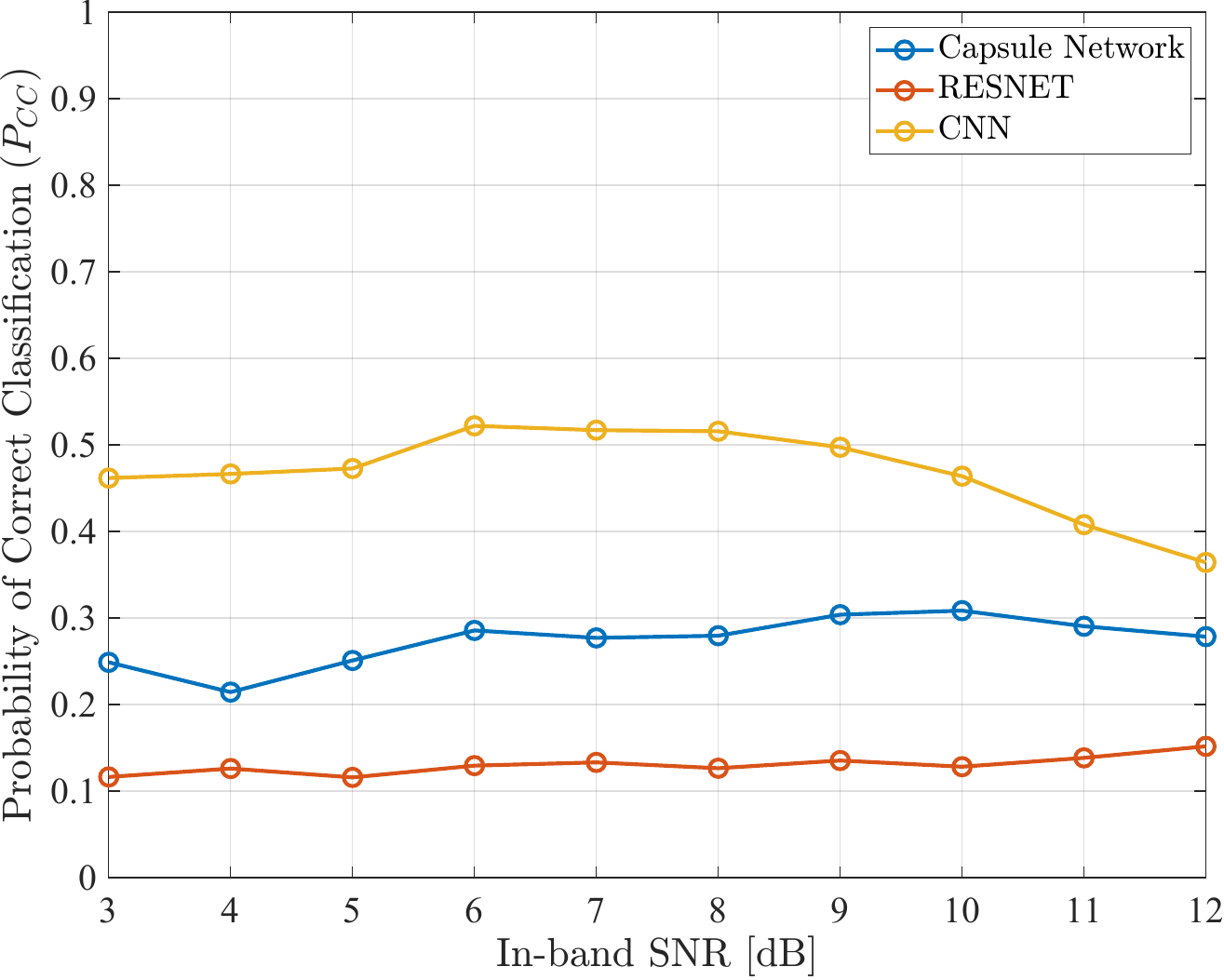}
\caption{Probability of correct classification versus SNR for training with \texttt{DataSet1} and testing
with \texttt{DataSet2}.}\label{fig:snr_gentest}
\end{figure}

\begin{figure}
\centering
\includegraphics[scale=0.35]{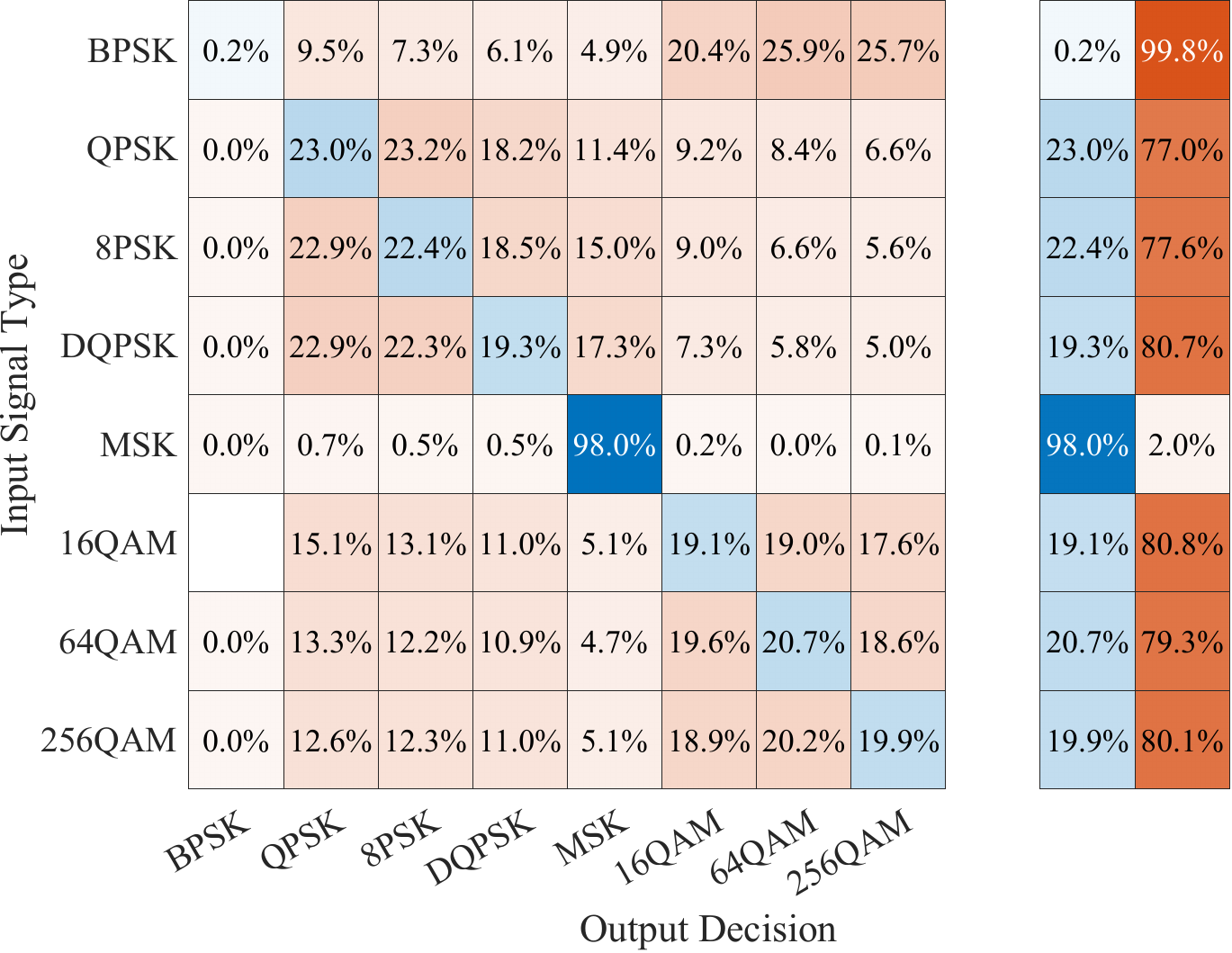}
\caption{Confusion matrix for the capsule network when training is done using \texttt{DataSet1} and testing uses data
from \texttt{DataSet2}.}\label{fig:cm_gentest}
\vspace{-0.5cm}
\end{figure}

Next, we studied the generalization ability of the capsule network, by using signals in \texttt{DataSet1} for training the network followed
by testing with signals in \texttt{DataSet2}. Results from this experiment are shown in Fig.~\ref{fig:snr_gentest} and Fig.~\ref{fig:cm_gentest}
from where we note that when the capsule network trained using signals in \texttt{DataSet1} was tested with signals in \texttt{DataSet2},
the classification performance degraded significantly to an overall probability of correct classification of only $27.925\%$ as can be
observed from Fig.~\ref{fig:snr_gentest}, indicating that the capsule network trained using \texttt{DataSet1} does not appear to
generalize training to a different dataset.


The lack of ability to generalize training is also observed if the capsule network is re-trained using signals from \texttt{DataSet2} and then
tested using signals in \texttt{DataSet1}, in which case a similar overall probability of correct classification of $26.2107\%$ is obtained.
Due to space constraints we omit the plots showing the variation of probability of correct classification vs. SNR and the confusion matrix
for this experiment, as these are similar to the ones shown in Fig.~\ref{fig:snr_gentest} and Fig.~\ref{fig:cm_gentest}, respectively. Thus,
we conclude that, while the capsule network is able to adapt well to the dataset change and re-learn to correctly classify modulation types
in a new dataset with high accuracy, it is not able to generalize its training to maintain good classification performance when presented with
signals in a different dataset. Nevertheless, we note that the capsule network does appear to learn some baseline signal features that are
common to both \texttt{DataSet1} and \texttt{DataSet2}, as the overall probability of correct classification in both cases is more than double
that of a random guess\footnote{With $8$ digital modulation schemes to be classified there is a $12.5\%$ chance of a random guess being
correct.}, but that the classification performance is sensitive to the different CFOs or symbol rates in the two datasets.


\begin{figure}
\centering
\includegraphics[scale=0.35]{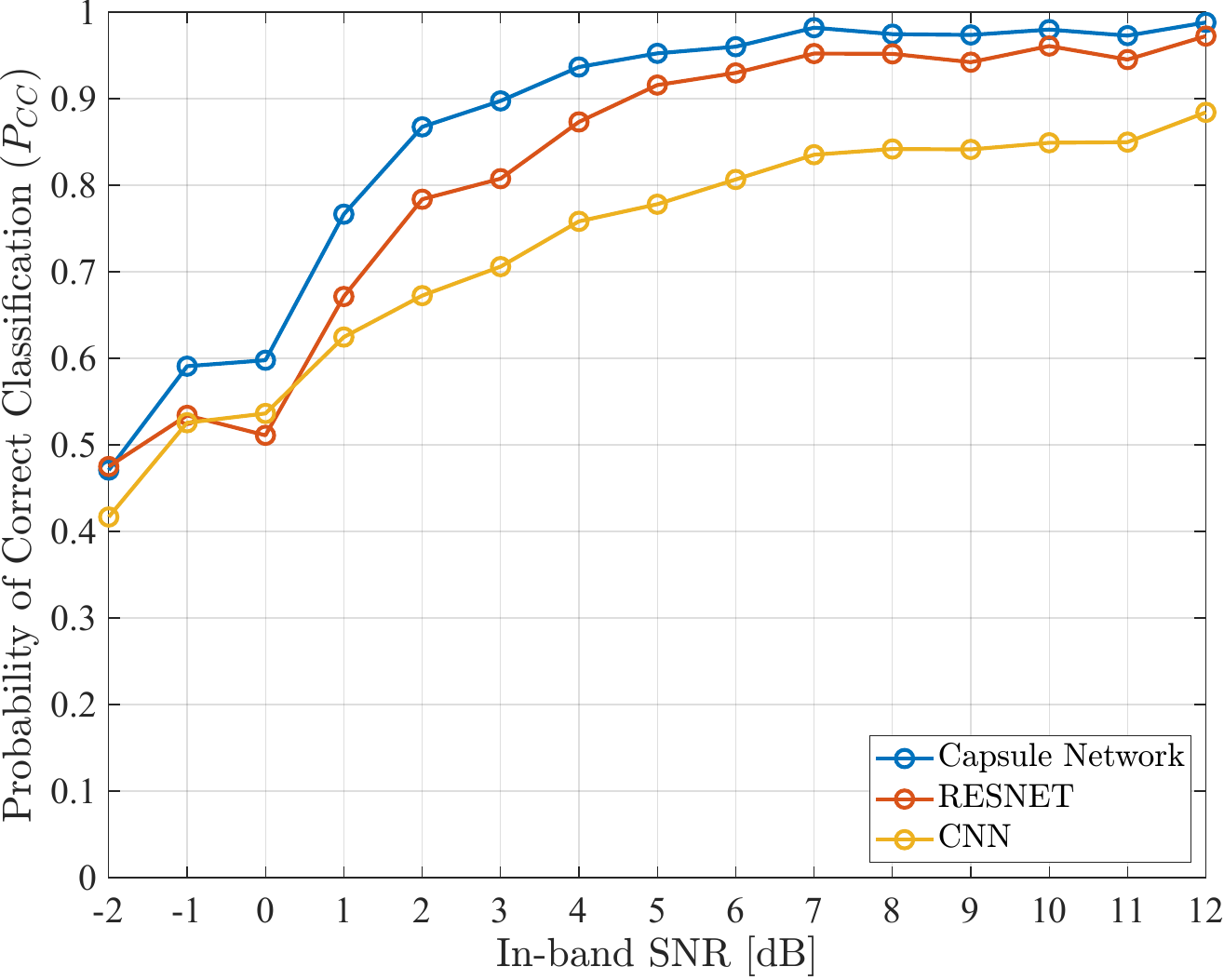}
\caption{Probability of correct classification versus SNR when both training and test data come from a mix
of \texttt{DataSet1} and \texttt{DataSet2}.}\label{fig:snr_gc_c}
\end{figure}

\begin{figure}
\includegraphics[scale=0.35]{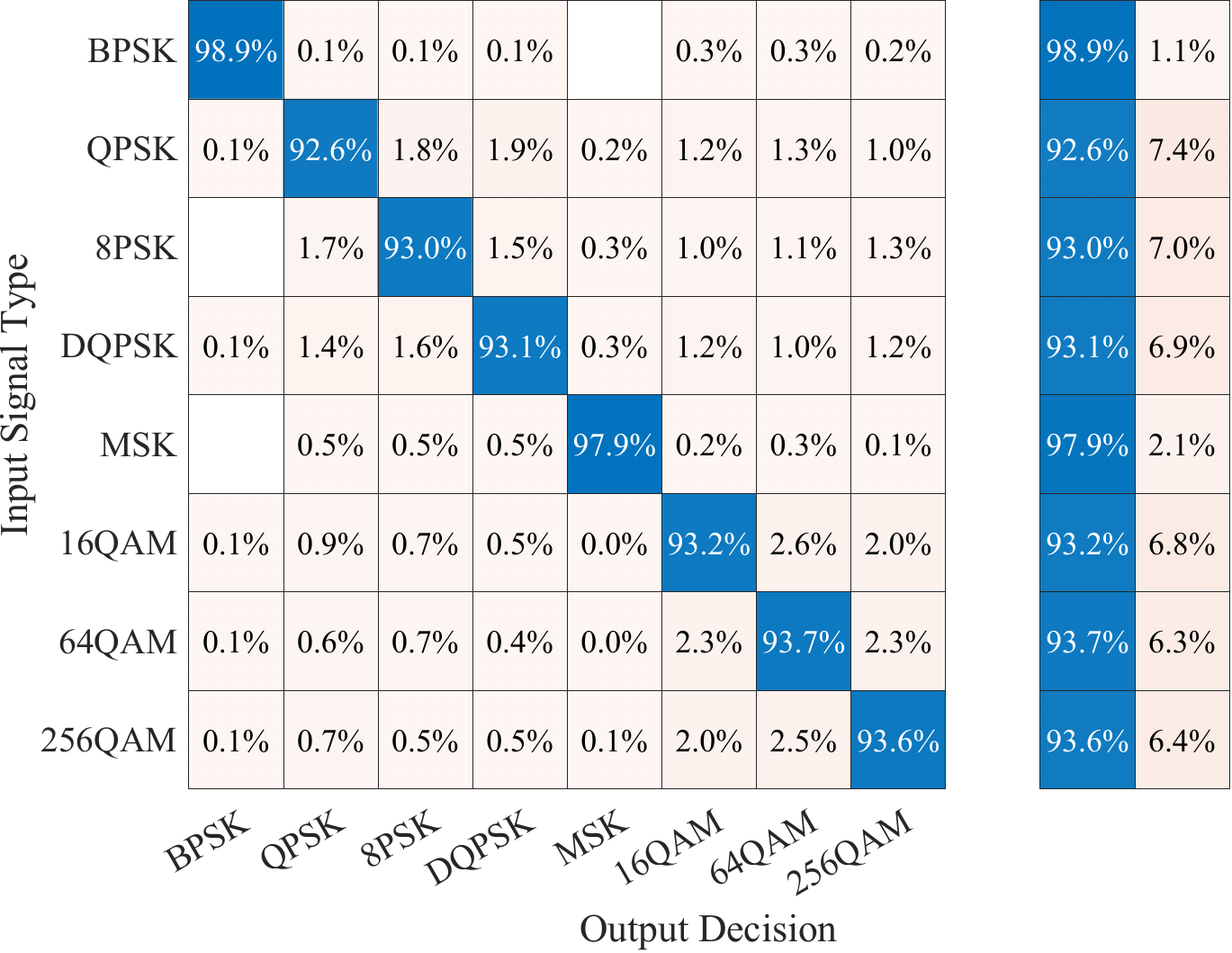}
\caption{Confusion matrix when both training and test data come from a mix of \texttt{DataSet1} and \texttt{DataSet2}.}\label{fig:cm_gc_c}
\vspace{-0.45cm}
\end{figure}

In a final experiment, we combined the signals in \texttt{DataSet1} and \texttt{DataSet2} and trained the capsule network with this new,
mixed dataset. We note that, while a total of $224,000$ signals are available in the two datasets, due to storage and memory constraints
imposed by the high-performance cluster hardware and operation, we included only $160,000$ signals in the combined dataset, randomly
taking $80,000$ signals from \texttt{DataSet1} and  $80,0000$ from \texttt{DataSet2} to make up the mixed dataset containing $160,000$
digitally modulated signals. Following a similar approach as for previous experiments, the $160,000$ signals in the combined dataset were
divided into three categories, with $70\%$ of signals used for training, $5\%$ for validation, and the remaining $25\%$ of the signals used
for testing. The results of this experiment are shown in Fig.~\ref{fig:snr_gc_c} and Fig.~\ref{fig:cm_gc_c}, from which we note an overall
performance of $94.4975\%$ correct classification, with probabilities of correct classification of individual modulation schemes ranging
from $92.6\%$ for QPSK to $98.9\%$ for BPSK.


\section{Conclusions}\label{sec:Conclusion}
This paper explored the use of capsule networks for classification of digitally modulated signals using the raw I/Q components of the
modulated signal. The overall classification performance implied by capsule networks is on a par or exceeds that obtained in related
work where CNNs or residual networks (RESNETs) are used \cite{Tim2018, Snoap_CCNC2022}, indicating that capsule networks are a
meaningful alternative for machine learning approaches to digitally modulated signal classification. We note that, similar to CNNs
and RESNETs, when trained with the raw I/Q signal data, capsule networks are able to learn characteristics of the signals in the dataset
used for training, but they are not able to generalize their learning to new datasets, which contain similar types of digitally modulated
signals but with differences in some of their characteristics such as the CFO or symbol period.

To overcome the problem of generalization, the capsule network was also trained with a mix of datasets, which provided additional
training data that contains desired learnable characteristics. This approach improved overall performance of the capsule network when
tested with signals from the mixed datasets, acknowledging the expectation that given sufficient training data the overall classification
performance improves. To further improve the generalization ability of capsule networks in classifying digitally modulated signals,
future work will consider training them using specific features of digitally modulated signals that can be extracted from the raw I/Q
signal data, such as those based on cyclostationary signal processing.

\section*{Acknowledgment}
The authors would like to acknowledge the use of Old Dominion University High-Performance Computing facilities
for obtaining numerical results presented in this work.


\IEEEtriggeratref{6}

%
%
 \bibliographystyle{IEEEtran}
 \bibliography{comm2022_arXiv}
%
%
%

\end{document}